\documentclass[conference,10pt]{IEEEtran}
\IEEEoverridecommandlockouts

\usepackage{cite}
\usepackage{amsmath,amsfonts,amssymb,amsthm}
\usepackage{algorithm}
\usepackage{algorithmic}
\usepackage{graphicx}
\usepackage{textcomp}
\usepackage{xcolor}
\usepackage{microtype}
\usepackage{balance}
\usepackage[caption=false,font=footnotesize]{subfig}

\newcommand{\eg}[1]{\textcolor{green}{\bf[eg: #1]}}

\allowdisplaybreaks 
\makeatletter
\newcommand\fs@betterruled{%
  \def\@fs@cfont{\bfseries}\let\@fs@capt\floatc@ruled
  \def\@fs@pre{\vspace*{5pt}\hrule height.8pt depth0pt \kern2pt}%
  \def\@fs@post{\kern2pt\hrule\relax}%
  \def\@fs@mid{\kern2pt\hrule\kern2pt}%
  \let\@fs@iftopcapt\iftrue}
\floatstyle{betterruled}
\restylefloat{algorithm}
\makeatother

\newtheorem{definition}{Definition}
\usepackage{amssymb}
\usepackage{amsfonts}
\usepackage{mathrsfs}
\usepackage{xspace}
\usepackage{bm}
\usepackage{upgreek}

\newcommand{\safemath}[2]{\newcommand{#1}{\ensuremath{#2}\xspace}}

\safemath{\bma}{\mathbf{a}}
\safemath{\bmb}{\mathbf{b}}
\safemath{\bmc}{\mathbf{c}}
\safemath{\bmd}{\mathbf{d}}
\safemath{\bme}{\mathbf{e}}
\safemath{\bmf}{\mathbf{f}}
\safemath{\bmg}{\mathbf{g}}
\safemath{\bmh}{\mathbf{h}}
\safemath{\bmi}{\mathbf{i}}
\safemath{\bmj}{\mathbf{j}}
\safemath{\bmk}{\mathbf{k}}
\safemath{\bml}{\mathbf{l}}
\safemath{\bmm}{\mathbf{m}}
\safemath{\bmn}{\mathbf{n}}
\safemath{\bmo}{\mathbf{o}}
\safemath{\bmp}{\mathbf{p}}
\safemath{\bmq}{\mathbf{q}}
\safemath{\bmr}{\mathbf{r}}
\safemath{\bms}{\mathbf{s}}
\safemath{\bmt}{\mathbf{t}}
\safemath{\bmu}{\mathbf{u}}
\safemath{\bmv}{\mathbf{v}}
\safemath{\bmw}{\mathbf{w}}
\safemath{\bmx}{\mathbf{x}}
\safemath{\bmy}{\mathbf{y}}
\safemath{\bmz}{\mathbf{z}}
\safemath{\bmzero}{\mathbf{0}}
\safemath{\bmone}{\mathbf{1}}

\bmdefine{\biad}{a}
\bmdefine{\bibd}{b}
\bmdefine{\bicd}{c}
\bmdefine{\bidd}{d}
\bmdefine{\bied}{e}
\bmdefine{\bifd}{f}
\bmdefine{\bigd}{g}
\bmdefine{\bihd}{h}
\bmdefine{\biid}{i}
\bmdefine{\bijd}{j}
\bmdefine{\bikd}{k}
\bmdefine{\bild}{l}
\bmdefine{\bimd}{m}
\bmdefine{\bind}{n}
\bmdefine{\biod}{o}
\bmdefine{\bipd}{p}
\bmdefine{\biqd}{q}
\bmdefine{\bird}{r}
\bmdefine{\bisd}{s}
\bmdefine{\bitd}{t}
\bmdefine{\biud}{u}
\bmdefine{\bivd}{v}
\bmdefine{\biwd}{w}
\bmdefine{\bixd}{x}
\bmdefine{\biyd}{y}
\bmdefine{\bizd}{z}

\bmdefine{\bixid}{\xi}
\bmdefine{\bilambdad}{\lambda}
\bmdefine{\bimud}{\mu}
\bmdefine{\bithetad}{\theta}
\bmdefine{\biphid}{\phi}
\bmdefine{\bideltad}{\delta}

\safemath{\bmia}{\biad}
\safemath{\bmib}{\bibd}
\safemath{\bmic}{\bicd}
\safemath{\bmid}{\bidd}
\safemath{\bmie}{\bied}
\safemath{\bmif}{\bifd}
\safemath{\bmig}{\bigd}
\safemath{\bmih}{\bihd}
\safemath{\bmii}{\biid}
\safemath{\bmij}{\bijd}
\safemath{\bmik}{\bikd}
\safemath{\bmil}{\bild}
\safemath{\bmim}{\bimd}
\safemath{\bmin}{\bind}
\safemath{\bmio}{\biod}
\safemath{\bmip}{\bipd}
\safemath{\bmiq}{\biqd}
\safemath{\bmir}{\bird}
\safemath{\bmis}{\bisd}
\safemath{\bmit}{\bitd}
\safemath{\bmiu}{\biud}
\safemath{\bmiv}{\bivd}
\safemath{\bmiw}{\biwd}
\safemath{\bmix}{\bixd}
\safemath{\bmiy}{\biyd}
\safemath{\bmiz}{\bizd}

\safemath{\bmxi}{\bixid}
\safemath{\bmlambda}{\bilambdad}
\safemath{\bmmu}{\bimud}
\safemath{\bmtheta}{\bithetad}
\safemath{\bmphi}{\biphid}
\safemath{\bmdelta}{\bideltad}

\safemath{\bA}{\mathbf{A}}
\safemath{\bB}{\mathbf{B}}
\safemath{\bC}{\mathbf{C}}
\safemath{\bD}{\mathbf{D}}
\safemath{\bE}{\mathbf{E}}
\safemath{\bF}{\mathbf{F}}
\safemath{\bG}{\mathbf{G}}
\safemath{\bH}{\mathbf{H}}
\safemath{\bI}{\mathbf{I}}
\safemath{\bJ}{\mathbf{J}}
\safemath{\bK}{\mathbf{K}}
\safemath{\bL}{\mathbf{L}}
\safemath{\bM}{\mathbf{M}}
\safemath{\bN}{\mathbf{N}}
\safemath{\bO}{\mathbf{O}}
\safemath{\bP}{\mathbf{P}}
\safemath{\bQ}{\mathbf{Q}}
\safemath{\bR}{\mathbf{R}}
\safemath{\bS}{\mathbf{S}}
\safemath{\bT}{\mathbf{T}}
\safemath{\bU}{\mathbf{U}}
\safemath{\bV}{\mathbf{V}}
\safemath{\bW}{\mathbf{W}}
\safemath{\bX}{\mathbf{X}}
\safemath{\bY}{\mathbf{Y}}
\safemath{\bZ}{\mathbf{Z}}

\safemath{\bZero}{\mathbf{0}}
\safemath{\bOne}{\mathbf{1}}
\safemath{\bDelta}{\mathbf{\Delta}}
\safemath{\bLambda}{\mathbf{\UpLambda}}
\safemath{\bPhi}{\mathbf{\Phi}}
\safemath{\bPsi}{\mathbf{\Psi}}
\safemath{\bSigma}{\mathbf{\Upsigma}}
\safemath{\bOmega}{\mathbf{\Upomega}}
\safemath{\bTheta}{\mathbf{\Uptheta}}

\bmdefine{\biAd}{A}
\bmdefine{\biBd}{B}
\bmdefine{\biCd}{C}
\bmdefine{\biDd}{D}
\bmdefine{\biEd}{E}
\bmdefine{\biFd}{F}
\bmdefine{\biGd}{G}
\bmdefine{\biHd}{H}
\bmdefine{\biId}{I}
\bmdefine{\biJd}{J}
\bmdefine{\biKd}{K}
\bmdefine{\biLd}{L}
\bmdefine{\biMd}{M}
\bmdefine{\biOd}{N}
\bmdefine{\biPd}{O}
\bmdefine{\biQd}{P}
\bmdefine{\biRd}{R}
\bmdefine{\biSd}{S}
\bmdefine{\biTd}{T}
\bmdefine{\biUd}{U}
\bmdefine{\biVd}{V}
\bmdefine{\biWd}{W}
\bmdefine{\biXd}{X}
\bmdefine{\biYd}{Y}
\bmdefine{\biZd}{Z}

\bmdefine{\biDelta}{\Delta}
\bmdefine{\biLambda}{\Lambda}
\bmdefine{\biPhi}{\Phi}
\bmdefine{\biSigma}{\Sigma}
\bmdefine{\biOmega}{\Omega}
\bmdefine{\biTheta}{\Theta}

\safemath{\bimA}{\biAd}
\safemath{\bimB}{\biBd}
\safemath{\bimC}{\biCd}
\safemath{\bimD}{\biDd}
\safemath{\bimE}{\biEd}
\safemath{\bimF}{\biFd}
\safemath{\bimG}{\biGd}
\safemath{\bimH}{\biHd}
\safemath{\bimI}{\biId}
\safemath{\bimJ}{\biJd}
\safemath{\bimK}{\biKd}
\safemath{\bimL}{\biLd}
\safemath{\bimM}{\biMd}
\safemath{\bimN}{\biNd}
\safemath{\bimO}{\biOd}
\safemath{\bimP}{\biPd}
\safemath{\bimQ}{\biQd}
\safemath{\bimR}{\biRd}
\safemath{\bimS}{\biSd}
\safemath{\bimT}{\biTd}
\safemath{\bimU}{\biUd}
\safemath{\bimV}{\biVd}
\safemath{\bimW}{\biWd}
\safemath{\bimX}{\biXd}
\safemath{\bimY}{\biYd}
\safemath{\bimZ}{\biZd}

\safemath{\bimDelta}{\biDelta}
\safemath{\bimLambda}{\biLambda}
\safemath{\bimPhi}{\biPhi}
\safemath{\bimSigma}{\biSigma}
\safemath{\bimOmega}{\biOmega}
\safemath{\bimTheta}{\biTheta}

\safemath{\setA}{\mathcal{A}}
\safemath{\setB}{\mathcal{B}}
\safemath{\setC}{\mathcal{C}}
\safemath{\setD}{\mathcal{D}}
\safemath{\setE}{\mathcal{E}}
\safemath{\setF}{\mathcal{F}}
\safemath{\setG}{\mathcal{G}}
\safemath{\setH}{\mathcal{H}}
\safemath{\setI}{\mathcal{I}}
\safemath{\setJ}{\mathcal{J}}
\safemath{\setK}{\mathcal{K}}
\safemath{\setL}{\mathcal{L}}
\safemath{\setM}{\mathcal{M}}
\safemath{\setN}{\mathcal{N}}
\safemath{\setO}{\mathcal{O}}
\safemath{\setP}{\mathcal{P}}
\safemath{\setQ}{\mathcal{Q}}
\safemath{\setR}{\mathcal{R}}
\safemath{\setS}{\mathcal{S}}
\safemath{\setT}{\mathcal{T}}
\safemath{\setU}{\mathcal{U}}
\safemath{\setV}{\mathcal{V}}
\safemath{\setW}{\mathcal{W}}
\safemath{\setX}{\mathcal{X}}
\safemath{\setY}{\mathcal{Y}}
\safemath{\setZ}{\mathcal{Z}}
\safemath{\emptySet}{\varnothing}

\safemath{\colA}{\mathscr{A}}
\safemath{\colB}{\mathscr{B}}
\safemath{\colC}{\mathscr{C}}
\safemath{\colD}{\mathscr{D}}
\safemath{\colE}{\mathscr{E}}
\safemath{\colF}{\mathscr{F}}
\safemath{\colG}{\mathscr{G}}
\safemath{\colH}{\mathscr{H}}
\safemath{\colI}{\mathscr{I}}
\safemath{\colJ}{\mathscr{J}}
\safemath{\colK}{\mathscr{K}}
\safemath{\colL}{\mathscr{L}}
\safemath{\colM}{\mathscr{M}}
\safemath{\colN}{\mathscr{N}}
\safemath{\colO}{\mathscr{O}}
\safemath{\colP}{\mathscr{P}}
\safemath{\colQ}{\mathscr{Q}}
\safemath{\colR}{\mathscr{R}}
\safemath{\colS}{\mathscr{S}}
\safemath{\colT}{\mathscr{T}}
\safemath{\colU}{\mathscr{U}}
\safemath{\colV}{\mathscr{V}}
\safemath{\colW}{\mathscr{W}}
\safemath{\colX}{\mathscr{X}}
\safemath{\colY}{\mathscr{Y}}
\safemath{\colZ}{\mathscr{Z}}

\safemath{\opA}{\mathbb{A}}
\safemath{\opB}{\mathbb{B}}
\safemath{\opC}{\mathbb{C}}
\safemath{\opD}{\mathbb{D}}
\safemath{\opE}{\mathbb{E}}
\safemath{\opF}{\mathbb{F}}
\safemath{\opG}{\mathbb{G}}
\safemath{\opH}{\mathbb{H}}
\safemath{\opI}{\mathbb{I}}
\safemath{\opJ}{\mathbb{J}}
\safemath{\opK}{\mathbb{K}}
\safemath{\opL}{\mathbb{L}}
\safemath{\opM}{\mathbb{M}}
\safemath{\opN}{\mathbb{N}}
\safemath{\opO}{\mathbb{O}}
\safemath{\opP}{\mathbb{P}}
\safemath{\opQ}{\mathbb{Q}}
\safemath{\opR}{\mathbb{R}}
\safemath{\opS}{\mathbb{S}}
\safemath{\opT}{\mathbb{T}}
\safemath{\opU}{\mathbb{U}}
\safemath{\opV}{\mathbb{V}}
\safemath{\opW}{\mathbb{W}}
\safemath{\opX}{\mathbb{X}}
\safemath{\opY}{\mathbb{Y}}
\safemath{\opZ}{\mathbb{Z}}
\safemath{\opZero}{\mathbb{O}}
\safemath{\identityop}{\opI}

\safemath{\veca}{\bma}
\safemath{\vecb}{\bmb}
\safemath{\vecc}{\bmc}
\safemath{\vecd}{\bmd}
\safemath{\vece}{\bme}
\safemath{\vecf}{\bmf}
\safemath{\vecg}{\bmg}
\safemath{\vech}{\bmh}
\safemath{\veci}{\bmi}
\safemath{\vecj}{\bmj}
\safemath{\veck}{\bmk}
\safemath{\vecl}{\bml}
\safemath{\vecm}{\bmm}
\safemath{\vecn}{\bmn}
\safemath{\veco}{\bmo}
\safemath{\vecp}{\bmp}
\safemath{\vecq}{\bmq}
\safemath{\vecr}{\bmr}
\safemath{\vecs}{\bms}
\safemath{\vect}{\bmt}
\safemath{\vecu}{\bmu}
\safemath{\vecv}{\bmv}
\safemath{\vecw}{\bmw}
\safemath{\vecx}{\bmx}
\safemath{\vecy}{\bmy}
\safemath{\vecz}{\bmz}

\safemath{\veczero}{\bmzero}
\safemath{\vecone}{\bmone}
\safemath{\vecxi}{\bmxi}
\safemath{\veclambda}{\bmlambda}
\safemath{\vecmu}{\bmmu}
\safemath{\vectheta}{\bmtheta}
\safemath{\vecphi}{\bmphi}
\safemath{\vecdelta}{\bmdelta}

\safemath{\matA}{\bA}
\safemath{\matB}{\bB}
\safemath{\matC}{\bC}
\safemath{\matD}{\bD}
\safemath{\matE}{\bE}
\safemath{\matF}{\bF}
\safemath{\matG}{\bG}
\safemath{\matH}{\bH}
\safemath{\matI}{\bI}
\safemath{\matJ}{\bJ}
\safemath{\matK}{\bK}
\safemath{\matL}{\bL}
\safemath{\matM}{\bM}
\safemath{\matN}{\bN}
\safemath{\matO}{\bO}
\safemath{\matP}{\bP}
\safemath{\matQ}{\bQ}
\safemath{\matR}{\bR}
\safemath{\matS}{\bS}
\safemath{\matT}{\bT}
\safemath{\matU}{\bU}
\safemath{\matV}{\bV}
\safemath{\matW}{\bW}
\safemath{\matX}{\bX}
\safemath{\matY}{\bY}
\safemath{\matZ}{\bZ}
\safemath{\matzero}{\bmzero}

\safemath{\matDelta}{\bDelta}
\safemath{\matLambda}{\bLambda}
\safemath{\matPhi}{\bPhi}
\safemath{\matSigma}{\bSigma}
\safemath{\matOmega}{\bOmega}
\safemath{\matTheta}{\bTheta}

\safemath{\matidentity}{\matI}
\safemath{\matone}{\matO}

\safemath{\rnda}{A}
\safemath{\rndb}{B}
\safemath{\rndc}{C}
\safemath{\rndd}{D}
\safemath{\rnde}{E}
\safemath{\rndf}{F}
\safemath{\rndg}{G}
\safemath{\rndh}{H}
\safemath{\rndi}{I}
\safemath{\rndj}{J}
\safemath{\rndk}{K}
\safemath{\rndl}{L}
\safemath{\rndm}{M}
\safemath{\rndn}{N}
\safemath{\rndo}{O}
\safemath{\rndp}{P}
\safemath{\rndq}{Q}
\safemath{\rndr}{R}
\safemath{\rnds}{S}
\safemath{\rndt}{T}
\safemath{\rndu}{U}
\safemath{\rndv}{V}
\safemath{\rndw}{W}
\safemath{\rndx}{X}
\safemath{\rndy}{Y}
\safemath{\rndz}{Z}

\safemath{\rveca}{\bimA}
\safemath{\rvecb}{\bimB}
\safemath{\rvecc}{\bimC}
\safemath{\rvecd}{\bimD}
\safemath{\rvece}{\bimE}
\safemath{\rvecf}{\bimF}
\safemath{\rvecg}{\bimG}
\safemath{\rvech}{\bimH}
\safemath{\rveci}{\bimI}
\safemath{\rvecj}{\bimJ}
\safemath{\rveck}{\bimK}
\safemath{\rvecl}{\bimL}
\safemath{\rvecm}{\bimM}
\safemath{\rvecn}{\bimN}
\safemath{\rveco}{\bomO}
\safemath{\rvecp}{\bimP}
\safemath{\rvecq}{\bimQ}
\safemath{\rvecr}{\bimR}
\safemath{\rvecs}{\bimS}
\safemath{\rvect}{\bimT}
\safemath{\rvecu}{\bimU}
\safemath{\rvecv}{\bimV}
\safemath{\rvecw}{\bimW}
\safemath{\rvecx}{\bimX}
\safemath{\rvecy}{\bimY}
\safemath{\rvecz}{\bimZ}

\safemath{\rvecxi}{\bmxi}
\safemath{\rveclambda}{\bmlambda}
\safemath{\rvecmu}{\bmmu}
\safemath{\rvectheta}{\bmtheta}
\safemath{\rvecphi}{\bmphi}

\safemath{\rmatA}{\bimA}
\safemath{\rmatB}{\bimB}
\safemath{\rmatC}{\bimC}
\safemath{\rmatD}{\bimD}
\safemath{\rmatE}{\bimE}
\safemath{\rmatF}{\bimF}
\safemath{\rmatG}{\bimG}
\safemath{\rmatH}{\bimH}
\safemath{\rmatI}{\bimI}
\safemath{\rmatJ}{\bimJ}
\safemath{\rmatK}{\bimK}
\safemath{\rmatL}{\bimL}
\safemath{\rmatM}{\bimM}
\safemath{\rmatN}{\bimN}
\safemath{\rmatO}{\bimO}
\safemath{\rmatP}{\bimP}
\safemath{\rmatQ}{\bimQ}
\safemath{\rmatR}{\bimR}
\safemath{\rmatS}{\bimS}
\safemath{\rmatT}{\bimT}
\safemath{\rmatU}{\bimU}
\safemath{\rmatV}{\bimV}
\safemath{\rmatW}{\bimW}
\safemath{\rmatX}{\bimX}
\safemath{\rmatY}{\bimY}
\safemath{\rmatZ}{\bimZ}

\safemath{\rmatDelta}{\bimDelta}
\safemath{\rmatLambda}{\bimLambda}
\safemath{\rmatPhi}{\bimPhi}
\safemath{\rmatSigma}{\bimSigma}
\safemath{\rmatOmega}{\bimOmega}
\safemath{\rmatTheta}{\bimTheta}

\usepackage{amssymb}
\usepackage{amsfonts}
\usepackage{mathrsfs}
\usepackage{xspace}
\usepackage{bm}
\usepackage{fancyref}
\usepackage{textcomp}

\usepackage{multirow}
\usepackage{stmaryrd}

\newenvironment{textbmatrix}{	\setlength{\arraycolsep}{2.5pt}%
								\big[\begin{matrix}}{\end{matrix}\big]%
								\raisebox{0.08ex}{\vphantom{M}}}

\def\be{\begin{equation}}
\def\ee{\end{equation}}
\def\een{\nonumber \end{equation}}
\def\mat{\begin{bmatrix}}
\def\emat{\end{bmatrix}}
\def\btm{\begin{textbmatrix}}
\def\etm{\end{textbmatrix}}

\def\ba#1\ea{\begin{align}#1\end{align}}
\def\bas#1\eas{\begin{align*}#1\end{align*}}
\def\bs#1\es{\begin{split}#1\end{split}} 
\def\bg#1\eg{\begin{gather}#1\end{gather}}
\def\bml#1\eml{\begin{multline}#1\end{multline}}
\def\bi#1\ei{\begin{itemize}#1\end{itemize}}

\DeclareMathOperator*{\argmin}{arg\;min}		
\DeclareMathOperator*{\argmax}{arg\;max}		

\safemath{\dirac}{\delta}					
\safemath{\krond}{\dirac}					

\safemath{\upto}{\uparrow}
\safemath{\downto}{\downarrow}
\safemath{\iu}{j}							
\safemath{\ev}{\lambda}						
\safemath{\hilseqspace}{l^{2}}				
\newcommand{\banachfunspace}[1]{\setL^{#1}}	
\safemath{\hilfunspace}{\banachfunspace{2}}	

\safemath{\SNR}{\text{\sc snr}} 				
\safemath{\No}{N_0}							
\safemath{\Es}{E_s}							
\safemath{\Eb}{E_b}							
\safemath{\EbNo}{\frac{\Eb}{\No}}
\safemath{\EsNo}{\frac{\Es}{\No}}

\DeclareMathOperator{\CHop}{\ensuremath{\opH}} 
\safemath{\tvir}{\rndh_{\CHop}}				
\safemath{\tvtf}{\rndl_{\CHop}}				
\safemath{\spf}{\rnds_{\CHop}}				
\safemath{\bff}{H_{\CHop}}					

\safemath{\ircf}{r_{h}}						
\safemath{\tftvcf}{r_{s}}					
\safemath{\tfcf}{r_{l}}						
\safemath{\bfcf}{r_{H}}						

\safemath{\tcorr}{c_h}						
\safemath{\scf}{c_{s}}						
\safemath{\tfcorr}{c_{l}}					
\safemath{\fcorr}{c_{H}}						

\safemath{\mi}{I}							
\safemath{\capacity}{C}						

\safemath{\normal}{\mathcal{N}}			
\safemath{\jpg}{\mathcal{CN}}			
\safemath{\mchain}{\leftrightarrow}		

\safemath{\dB}{\,\mathrm{dB}}
\safemath{\dBm}{\,\mathrm{dBm}}
\safemath{\Hz}{\,\mathrm{Hz}}
\safemath{\kHz}{\,\mathrm{kHz}}
\safemath{\MHz}{\,\mathrm{MHz}}
\safemath{\GHz}{\,\mathrm{GHz}}
\safemath{\s}{\,\mathrm{s}}
\safemath{\ms}{\,\mathrm{ms}}
\safemath{\mus}{\,\mathrm{\text{\textmu}s}}
\safemath{\ns}{\,\mathrm{ns}}
\safemath{\ps}{\,\mathrm{ps}}
\safemath{\meter}{\,\mathrm{m}}
\safemath{\mm}{\,\mathrm{mm}}
\safemath{\cm}{\,\mathrm{cm}}
\safemath{\W}{\,\mathrm{W}}
\safemath{\mW}{\, \mathrm{mW}}
\safemath{\J}{\,\mathrm{J}}
\safemath{\K}{\,\mathrm{K}}
\safemath{\bit}{\,\mathrm{bit}}
\safemath{\nat}{\,\mathrm{nat}}

\safemath{\define}{\triangleq}			

\safemath{\equivalent}{\sim}
\safemath{\distas}{\sim}					
\safemath{\sdiff}{\Delta}				

\safemath{\reals}{\mathbb{R}}
\safemath{\positivereals}{\reals_{+}}
\safemath{\integers}{\mathbb{Z}}
\safemath{\posint}{\integers_{+}}
\safemath{\naturals}{\mathbb{N}}
\safemath{\posnaturals}{\naturals_{+}}
\safemath{\complexset}{\mathbb{C}}
\safemath{\rationals}{\mathbb{Q}}

\newcommand*{\fancyrefapplabelprefix}{app}		
\newcommand*{\fancyrefthmlabelprefix}{thm}		
\newcommand*{\fancyreflemlabelprefix}{lem}		
\newcommand*{\fancyrefcorlabelprefix}{cor}		
\newcommand*{\fancyrefdeflabelprefix}{def}		
\newcommand*{\fancyrefproplabelprefix}{prop}		
\newcommand*{\fancyrefexmpllabelprefix}{exmpl}
\frefformat{vario}{\fancyrefseclabelprefix}{Section~#1}
\frefformat{vario}{\fancyrefthmlabelprefix}{Theorem~#1}
\frefformat{vario}{\fancyreflemlabelprefix}{Lemma~#1}
\frefformat{vario}{\fancyrefcorlabelprefix}{Corollary~#1}
\frefformat{vario}{\fancyrefdeflabelprefix}{Definition~#1}
\frefformat{vario}{\fancyreffiglabelprefix}{Fig.~#1}
\frefformat{vario}{\fancyrefapplabelprefix}{Appendix~#1}
\frefformat{vario}{\fancyrefeqlabelprefix}{(#1)}
\frefformat{vario}{\fancyrefproplabelprefix}{Property~#1}
\frefformat{vario}{\fancyrefexmpllabelprefix}{Example~#1}

\newtheorem{prop}{Proposition}

\begin{document}

\title{Identifying Unused RF Channels \\ Using Least  Matching Pursuit}

\author{\IEEEauthorblockN{Emre G\"{o}n\"{u}lta\c{s}, Milad Taghavi, Sweta Soni, Alyssa B. Apsel, and Christoph Studer}\\[-0.3cm]
	\IEEEauthorblockA{School of Electrical and Computer Engineering, Cornell University, Ithaca, NY
		 \\ e-mail: \{eg566,\,mt795,\,ss3964,\,aba25,\,studer\}@cornell.edu
		 }
	\thanks{The work of EG and CS was supported in part by Xilinx Inc.\ and by the US National Science Foundation (NSF) under grants CCF-1652065, CNS-1717559, and ECCS-1824379. MT, SS, and ABA were supported in part by the US NSF grant ECCS-1824379. The Quadro P6000 GPU used for this research was donated by the
		NVIDIA Corporation. }
}

\maketitle

\begin{abstract}
Cognitive radio aims at identifying unused radio-frequency (RF) bands with the goal of re-using them opportunistically for other services. 
While compressive sensing (CS) has been used to identify strong signals (or interferers) in the RF spectrum from sub-Nyquist measurements, identifying \emph{unused} frequencies from CS measurements appears to be uncharted territory.
In this paper, we propose a novel method for identifying unused RF bands using an algorithm we call least matching pursuit (LMP).
We present a sufficient condition for which LMP is guaranteed to identify unused frequency bands and develop an improved algorithm that is  inspired by our theoretical result. 
We perform simulations for a CS-based RF whitespace detection task in order to demonstrate that LMP is able to outperform black-box approaches that build  on deep neural networks. 
\end{abstract}

\section{Introduction}

In 2019, approximately 10.8 billion Internet of things (IoT) devices have been deployed worldwide, with the number of wirelessly connected devices growing at extreme rates over the last years~\cite{ericsson}. 
Without proper radio-frequency (RF) spectrum allocation strategies, the vast amount of wireless IoT devices would inevitably result in overcrowding of the available frequency resources. 
To optimally utilize the available spectrum, novel means to allocate IoT devices to unused frequencies are of paramount importance \cite{ghasemiiotunused}. 
While RF spectrum allocation can be performed at the infrastructure base station, identifying unused frequency bands must be performed at minimal power to reduce system costs~\cite{yucekspectrumsurvey}. Furthermore, enabling IoT devices with rudimentary whitespace detection capabilities would enable further improvements in terms of resource utilization as transmission could be scheduled opportunistically and more dynamically, when other nearby IoT transmitters are idle~\cite{zhangscheduling}. 
Consequently, both the infrastructure basestations and IoT devices would benefit from the development of novel means to identify unused frequencies in an energy-efficient manner~\cite{gaoiotspectrum}.

A straightforward way for detecting unused frequencies would be to sample the RF signal at Nyquist rate and analyze the spectrum in the Fourier domain~\cite{cabricspectrum}. 
To reduce the sampling rates and power consumption, a range of compressive sensing (CS)-based methods~\cite{4218361,6908362,6316045,6637036,a2f} have been proposed for spectrum sensing.
However, CS-based spectrum sensing is limited to detecting \emph{strong signals} and not designed for whitespace detection (i.e., identifying unused frequencies). In fact, CS  is typically unable to identify weak non-zero entries if they are only slightly above the noise floor~\cite{2012davenportnf}. 
As a sole exception, reference~\cite{yoo2013finding} proposed a method called   zero-detection group thresholding (ZD-GroTH), which is, to the best of our knowledge, the only CS-based method that has been designed to detect zero (unused) components.

\subsection{Contributions} 
In this paper, we improve upon the results of~\cite{yoo2013finding} in two ways: First, we develop a sufficient condition for which ZD-GroTH is guaranteed to identify zero (or unused) components; the condition depends on the dynamic range of the sparse signal, coherence properties of the sampling operator, and the noise power. 
Second, by inspecting our whitespace identification condition, we develop a refined ``anti-CS'' algorithm we call \emph{least matching pursuit} (LMP). Our method improves upon ZD-GroTH by including a block orthogonal matching pursuit (BOMP) stage~\cite{5424069,4960226}, which reduces the dynamic range between non-zero signal components, and a refined correlation criterion, which improves sensitivity. 
To demonstrate the efficacy of LMP, we simulate a whitespace detection task in a realistic RF system that measures spectral features using nonuniform wavelet sampling (NUWS)~\cite{8015153}, and we compare the performance of LMP to that of BOMP, ZF-GroTH, and a deep-learning based whitespace detector.

\subsection{Notation}
Uppercase boldface letters stand for matrices; lowercase boldface letters denote column vectors. 
For a matrix~$\bA$, we denote its Hermitian transpose by~$\bA^H$, its pseudo-inverse by~$\bA^\dagger$, and its $i$th block by $\bA_i$, which is a collection of contiguous columns in $\bA$. The $\ell_2$-norm (spectral norm) of a matrix~$\bA$ is $\|\bA\|_2=\sigma_{\max}$, where $\sigma_{\max}$ is the largest singular value of~$\bA$. 
The $\ell_2$-norm of a vector~$\bma$ is $\|\bma\|_2=\sqrt{\sum_{k}|a_k|^2}$.  

\section{System Model and Block-Sparse Recovery}
We now introduce the CS signal and measurement model, which we will use to model RF whitespace detection. 
Then, we summarize the recovery of non-zero blocks using BOMP and discuss its limitations for whitespace detection. 

\subsection{Compressive Sensing and Block-Sparse Signals}
To minimize the costs of sampling, we focus on CS-based acquisition schemes that acquire fewer measurements than the Nyquist rates dictates while exploiting signal sparsity in a given transform domain~\cite{1614066}. To model the fact that typical RF signals occupy frequency bands, we use a block-sparse signal model for which the signal components that are active or inactive appear in contiguous groups \cite{5424069,4960226}.

Mathematically, we assume a discretized signal $\bmx \in \mathbb{C}^{N}$ that consists of $B$ blocks $\bmx_i\in\complexset^{N_i}$, where $i=1,\ldots,B$, $\sum_{i=1}^B N_i = N$, and $\bmx=[\bmx_1^H,\ldots,\bmx_B^H]^H$.
For RF whitespace detection, each block $\bmx_i$, $i=1,\ldots,B$, represents a contiguous band of discrete frequencies. In what follows, we assume that the signal $\bmx$ is $K$-block-sparse, meaning that exactly $K$ blocks are nonzero. 
CS-based signal acquisition is modeled by the input-output relation  
\mbox{$\bmy=\bA \bmx + \bmn$},
where $\bmy \in \mathbb{C}^{M}$ contains the compressive measurements with $M \ll N$, $\bA \in \mathbb{C}^{M\times N}$ is the effective sensing matrix (which combines the effect of the sensing matrix and the sparsifying transform), and $\bmn \in \complexset^M$ models measurement noise. 
To simplify notation, we can write an equivalent system model $\bmy=\sum_{i \in \mathcal{U}} \bA_i \bmx_i +\bmn $, where $\bA_i\in\complexset^{M\times N_i}$ is the $i$th block matrix of the effective sensing matrix $\bA=[\bA_1,\ldots,\bA_B]$ and~$\mathcal{U}$ denotes the set of indices corresponding to the non-zero blocks in the vector $\bmx$. 

In practice, CS measurements are acquired by computing inner products between the uncompressed signal, written by $\bmz=\bPsi^{-1}\bmx$ where $\bPsi\in\complexset^{N\times N}$ is a sparsifying transform, and rows of the sensing matrix $\bTheta\in\complexset^{M\times N}$ using analog circuitry.  
The effective sensing matrix is $\bA=\bTheta\bPsi^{-1}$ and each block matrix is given by $\bA_i=\bTheta\bPsi^{-1}_i$, where $\bPsi^{-1}_i\in\complexset^{N\times N_i}$ is the $i$th block of the inverse sparsifying transform so that $\bPsi^{-1}=[\bPsi^{-1}_1,\ldots,\bPsi^{-1}_B]$.
For RF whitespace detection, $\bmz$ is the Nyquist-sampled time-domain signal, $\bPsi$ is the discrete Fourier transform (DFT) matrix (as we assume block-sparsity in the frequency domain), $\bTheta$ is a suitably-chosen sensing matrix (see \fref{sec:nuws} for the sensing matrix used in our experiments), and $\bmy$ contains the~$M$ compressive measurements 
\subsection{Block Orthogonal Matching Pursuit (BOMP)}

A prime goal of CS is to recover the nonzero (or used) blocks~$\bmx_i$, $i\in\setU$, in the vector $\bmx$ from the CS measurements in~$\bmy$ using block-sparse signal recovery algorithms~\cite{5424069}. In contrast, whitespace detection aims at detecting \emph{unused} blocks, i.e., the blocks indexed by the set $\setN = \{i=1,\ldots,N \mid \|\bmx_i\|_2=0\}$. 
For noiseless measurements, one could first run a block-sparse signal recovery algorithm to identify the used blocks and then label all other blocks as unused. The presence of noise, however, renders it extremely challenging to distinguish noise from weak signal components---the situation is further aggravated  by the fact that most (block) sparse signal recovery algorithms shrink weak signals to zero. 
Even though our goal is \emph{not} to detect strong signals, we now briefly summarize BOMP, a prominent block-sparse signal recovery method. As shown in \fref{sec:LMPsec}, we will use BOMP as a preprocessing step to improve CS-based whitespace detection.

BOMP is an iterative block-sparse signal recovery algorithm put forward in~\cite{4960226}. 
The algorithm starts by initializing the so-called residual by $\bmr^0=\bmy$. At each iteration $t=1,\ldots,K$, BOMP first calculates a normalized correlation between the residual $\bmr^t$ and each block $\bA_i,\, \forall  i\in \mathcal{B}$ according to 
\begin{align}
\lambda_i^{t+1}= \| (\bA_i^H\bA_i)^{-0.5}\bA_i^H \bmr^t\|_2.
\end{align}
Note that, compared to \cite{4960226}, we use a modified correlation criterion that uses the term $(\bA_i^H \bA_i)^{-0.5}$, which does not require the blocks to have orthonormal columns. 
Next, BOMP selects the index of the block with the highest correlation according to $c^{t+1}=\argmax_i \lambda_i^{t+1}$ and adds the selected block to the support set ${\Omega}^{t+1}={\Omega}^{t}\, \cup\,  {c^{t+1}}$. 
BOMP then computes an estimate~$\hat\bmx_{\Omega^{t+1}}$ of the non-zero blocks $\hat{\bmx}=\bA_{\Omega^{t+1}}^\dagger \bmy$, where~$\bA_{\Omega^{t+1}}$ contains the blocks indexed by the support set estimate~$\Omega^{t+1}$.
Finally, BOMP updates the residual according to $\bmr^{t+1}=\bmy-\bA_{\Omega^{t+1}} \hat{\bmx}_{\Omega^{t+1}}$. See Algorithm~\ref{alg:bomp} for the pseudocode of BOMP. 
\begin{algorithm}[tp]
	\caption{ Block Orthogonal Matching Pursuit (BOMP)}
	\label{alg:bomp}
	\begin{algorithmic}[1]
		\STATE  \textbf{input } $\{\bA_i\}_{i=1}^{B}$, $\bmy$, and $K$
		\STATE {\bf initialize } $\bmr^0=\bmy$ and $\Omega^0=\varnothing$
		\FOR{$t=1,\ldots,K$}
		\FOR{$i=1,\ldots,B$}
		\STATE $\lambda_i^{t+1}= \| (\bA_i^H\bA_i)^{-0.5}\bA_i^H \bmr^t\|_2$
		\ENDFOR
		\STATE{$c^{t+1}=\argmax_i \lambda_i^{t+1}$}
		\STATE{$\Omega^{t+1}=\Omega^{t}\, \cup \,c^{t+1}$}
		\STATE{$\hat{\bmx}_{\Omega^{t+1}}=\bA_{\Omega^{t+1}}^\dagger \bmy$}
		\STATE{$\bmr^{t+1}=\bmy-\bA_{\Omega^{t+1}}\hat{\bmx}_{\Omega^{t+1}}$}
		\ENDFOR
		\RETURN $\hat{\bmx}_{\Omega^{t+1}}$, $\Omega^{t+1}$, and $\{\lambda^{t+1}_i,i=1,\ldots,B\}_{t=1}^{K}$
	\end{algorithmic}
\end{algorithm}

\section{LMP: Least Matching Pursuit} \label{sec:LMPsec}
While (block) sparse signal recovery aims at recovering strong non-zero entries,  RF whitespace detection requires the identification of unused blocks. 
We now summarize the zero-detection group thresholding (ZD-GroTH) method from~\cite{yoo2013finding}, the only method we are aware of that has been proposed to detect unused signal components from CS measurements. We  provide a sufficient condition that guarantees successful detection of unused blocks in presence of noise. 
By using our theoretical result, we improve upon ZD-GroTH by including (i) a BOMP stage that reduces the dynamic range between non-zero signal components and (ii) a refined correlation criterion---we call the resulting method least matching pursuit (LMP). 

\subsection{Recovery Guarantee for ZD-GroTh}
The ZD-GroTh algorithm~\cite{yoo2013finding} identifies the block~$\bA_i$ that minimizes the correlation with the received signal as 
\begin{align}
\label{eq:antiomp}
\hat{f} = \argmin_{i=1,\ldots,B} \| (\bA_i^H\bA_i)^{-0.5}\bA_i^H\bmy\|_2.
\end{align}
In contrast to the original method in  \cite{yoo2013finding}, we use the selection criterion in \fref{eq:antiomp} that enables the use of unnormalized blocks. 
While \cite{yoo2013finding} provides statistical guarantees for the success of ZD-GroTh,  we next propose a sufficient condition for the success of ZD-GroTh that depends on the dynamic range of the block-sparse signal, the effective sensing matrix, and the power of the measurement noise.
In what follows, we will make use of the following definitions:
\begin{definition}
Let $\|\bmx_{\min}\|_2=\min_{j \in \mathcal{U}} \|\bmx_j\|_2$ be the $\ell_2$-norm of the block of $\bmx_i$ that has the minimum $\ell_2$-norm  among the used blocks indexed by $\mathcal{U}$. Let $\sigma_{\min }=\min_{i=1,\ldots,B} \sigma_{\bA_i}$ be the minimum singular value among all the blocks $\bA_i$, $i=1,\ldots,B$. Let the block mutual coherence of $\bA$ be 
\begin{align}
\label{eq:mu_block}
\mu_B = \max_{i\neq j} \|(\bA_i^H \bA_i)^{-0.5}\bA_i^H \bA_j\|_2.
\end{align}
\end{definition}

We can now formulate the following sufficient condition for the success of ZD-GroTh. The proof is given in \fref{app:bomp}.

\begin{prop}
\label{prop:bomp}
If the following condition holds	
\begin{align}
\label{eq:prop1}
\frac{\sum_{j \in \mathcal{U}} \|\bmx_j\|_2}{\|\bmx_{\min}\|_2} < \frac{1}{2} \left( \frac{\sigma_{\min }}{\mu_B}+1\right)- \frac{\|\bmn\|_2}{\mu_B \|\bmx_{\min}\|_2},
\end{align}
then ZD-GroTh \fref{eq:antiomp} is guaranteed to identify an unused block.
\end{prop}

Proposition 1 is useful to understand conditions for which we can identify an unused block via ZD-GroTh.
For the special case where (i) the blocks $\bA_i$, $i=1,\ldots,B$ have orthonormal columns, (ii)  all the used signal blocks have equal power, i.e., $\|\bmx_i\|=\|\bmx_j\|$ for $i\neq j$ and $i,j\in\setU$, (iii)  and for noiseless measurements, we recover the standard BOMP condition  $K< \frac{1}{2}\left( \mu_B^{-1}+1\right) $ from \cite{4960226}.
For the general case, it is clear that the dynamic range between strong signal components and the weakest one $\bmx_{\min}$ plays a critical role in the success of this method, i.e., the ratio $\delta = \frac{\sum_{j \in \mathcal{U}} \|\bmx_j\|_2}{\|\bmx_{\min}\|_2}$ on the LHS of \fref{eq:prop1} should be as small as possible\footnote{This ratio is lower-bounded by the block-sparsity $K$, which is achieved with equality if all active signal components have equal power.}. 
This observation enables us to design the following improved algorithm for detecting unused blocks from CS measurements.

\subsection{LMP: Least Matching Pursuit}
Inspired by the above observation, we minimize the detrimental effect of high dynamic range $\delta$ of active signal components by first eliminating the strongest active sparse blocks and then invoke a variant of the minimum correlation condition in~\fref{eq:antiomp}; this can be accomplished by first running BOMP for~$P$ iterations (typically $P\leq K$) followed by selecting the least correlated block from the resulting residual~$\bmr^{P+1}$. 
However, by directly using \fref{eq:antiomp} on the residual, we ignore correlation information that has been acquired during all BOMP iterations. 
To this end, we also refine the selection criterion by considering the sum of all correlation coefficients over the $P$ BOMP iterations and select the block with the smallest sum. 
This approach effectively avoids the selection of blocks that had small correlation only in the last iteration of BOMP but had consistently low correlation before. 
Concretely, we propose to use the following refined selection criterion:
\begin{align}
\hat{f} = \argmin_{i \in\{1,2,\ldots,B\}\backslash \Omega^{P+1}} \sum_{t=1}^P \lambda^{t+1}_i.
\end{align}
Here, the correlation results  $\{\lambda^{t+1}_i,i=1,\ldots,B\}_{t=1}^{P}$ have been collected while running BOMP as detailed in Algorithm~\ref{alg:bomp}.
The resulting whitespace detection method, called least matching pursuit (LMP), is summarized in Algorithm \ref{alg:lbmp}.

\begin{algorithm}[tp]
	\caption{Least  Matching Pursuit (LMP)}
	\label{alg:lbmp}
	\begin{algorithmic}[1]
		\STATE  \textbf{input } $\{\bA_i\}_{i=1}^{B}$, $\bmy$, and $P$ 
		\STATE Run BOMP for $P$ iterations to obtain $\Omega^{P+1}$ and $\{\lambda^{t+1}\}_{t=1}^P$
		\RETURN $\hat{f} = \argmin_{i \in\{1,2,\ldots,B\}\backslash \Omega^{P+1}} \sum_{t=1}^P \lambda^{t+1}_i$
		
	\end{algorithmic}
\end{algorithm}

\section{Results}
We now show simulation results for LMP in a CS-based whitespace detection task. We first detail the used compressive sensing strategy. We then detail the simulation setup. We finally show performance results for LMP compared to  BOMP, ZD-GroTH, and a deep-learning based whitespace detector. 

\subsection{Non-uniform Wavelet Sampling (NUWS)} \label{sec:nuws}
Non-uniform wavelet sampling (NUWS) has been proposed in~\cite{8015153} as a flexible and hardware-friendly compressive sensing strategy for RF sensing~\cite{michaeliscas} and feature extraction tasks~\cite{eusipco_a2f}. In short, NUWS combines the advantages of nonuniform sampling~\cite{6316045} and random modulation~\cite{4472247}. 
Mathematically, NUWS takes inner products between the analog signal and wavelet-like pulses $\bmw(\tau,\rho,f) \in \mathbb{C}^{N}$, which can be tuned in terms of the time instant $\tau$, pulse width $\rho$, and frequency $f$. 
The tunability of NUWS enables one to adapt the sequence of wavelets to the task at hand. For our RF sensing application, we consider Haar-like wavelets~\cite{eusipco_a2f} with entries in $\{+1,0,-1\}$ that can easily be generated with mixed-signal circuitry~\cite{michaeliscas}. 

In order to adapt the sensing matrix to our application, we first construct an overcomplete dictionary $\bW\in\complexset^{L\times N}$ whose rows consists of a large number of different wavelet sequences $\{\bmw(\tau_l,\rho_l,f_l)\}_{l=1}^L$ with $L\gg N$. We then select a subset of~$M$ wavelets so that the effective sensing matrix $\bA=\bR_\Omega \bW$ has desirable properties. Here, $\bR_\Omega$ is a restriction operator that selects a subset of $M$ sequences from the dictionary~$\bW$.
Subset selection is carried out so that the resulting block mutual coherence $\mu_B$ in \fref{eq:mu_block} is minimized. Since this subset selection problem is of combinatorial nature, we use a greedy approach.
We start with an empty set of wavelet sequences. We then calculate the block mutual coherence $\mu_B$ for each of the $l= 1, \ldots, L$ wavelet sequences in $\bW$ and keep the sequence associated with the lowest $\mu_B$. We repeat this greedy procedure until $M$ wavelets have been collected. 

\begin{figure}[tp]
\centering
\includegraphics[width=0.9\columnwidth]{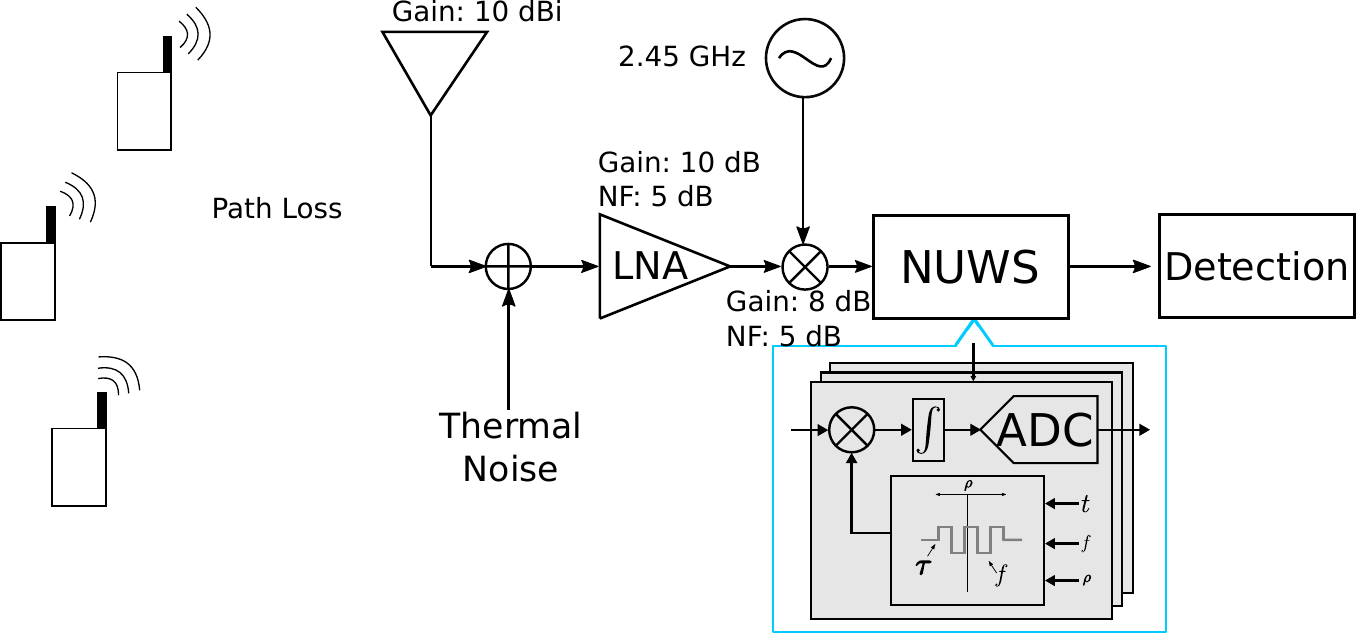}
\vspace{-0.1cm}
\caption{Block diagram of the system model representing RF transmitters and an RF receiver that performs NUWS and whitespace detection.}	
\label{fig:model}
\end{figure}

\subsection{Simulation Setup}

To demonstrate the efficacy of LMP in a realistic scenario, we model a complete RF transmitter and receiver chain in MATLAB; see \fref{fig:model} for an overview. We generate $20$ frequency bands between $2.4$\,GHz and $2.5$\,GHz and assume that at most five transmitters in five bands are active at a time. 
We randomly place the transmitters within a distance of $1$\,m and $280$\,m from the receiver and use the path-loss model from~\cite{944859}. 
At the transmitter, we generate a $5$\,MHz QPSK signal that is mixed with a carrier that is randomly selected among the 20 uniformly spaced channels. The $20$\,dBm signal is then transmitted over an antenna at height $1.65$\,m. 

At the receiver side, we collect the signals at an antenna at height $15$\,m with an antenna gain of $10$\,dBi.
The signal is then passed through a low-noise amplifier (LNA) followed by a intermediate frequency mixer operating at $2.45$\,GHz.
We model LNA and mixed non-linearities using first, third, and fifth harmonics at $50$\,Ohm impedance, $-1$\,dB gain compression, and a  third-order intercept point of $10$\,dBm. Phase noise at the mixers is simulated using Leeson's model \cite{1446612} with $1$\,MHz carrier frequency offset at $-110$\,dBc. 
For the voltage gains of mixers and the LNA, we use $8$\,dB and $20$\,dB respectively, and we use a noise figure of $5$\,dB for both the mixer and LNA. We add thermal noise to the signal both at the transmitter before the mixer and at the receiver before LNA operating at $290$\,K.
We then perform NUWS on the output signal of the RF receiver using the wavelet dictionary described in  \fref{sec:nuws} by taking a maximum of $N=200$ Nyquist samples.

\newcommand{\figsize}{0.32}
\begin{figure*}[tp]
	\center
	\label{fig:results}	
	\subfloat[50 NUWS Samples]
	{
		\includegraphics[width=\figsize\textwidth]{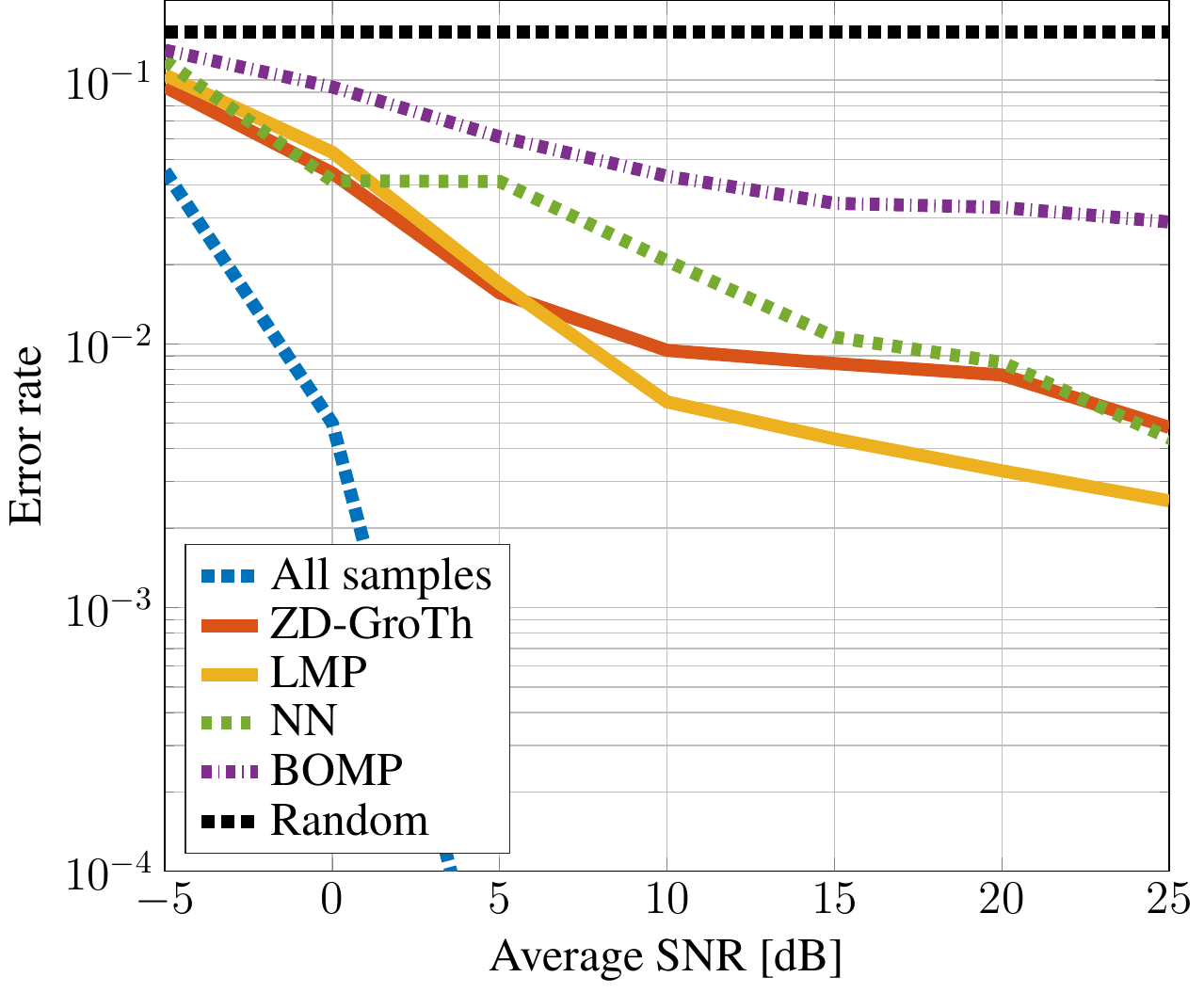}
		\label{fig:50}
	}
	\subfloat[100 NUWS Samples]
	{
		\includegraphics[width=\figsize\textwidth]{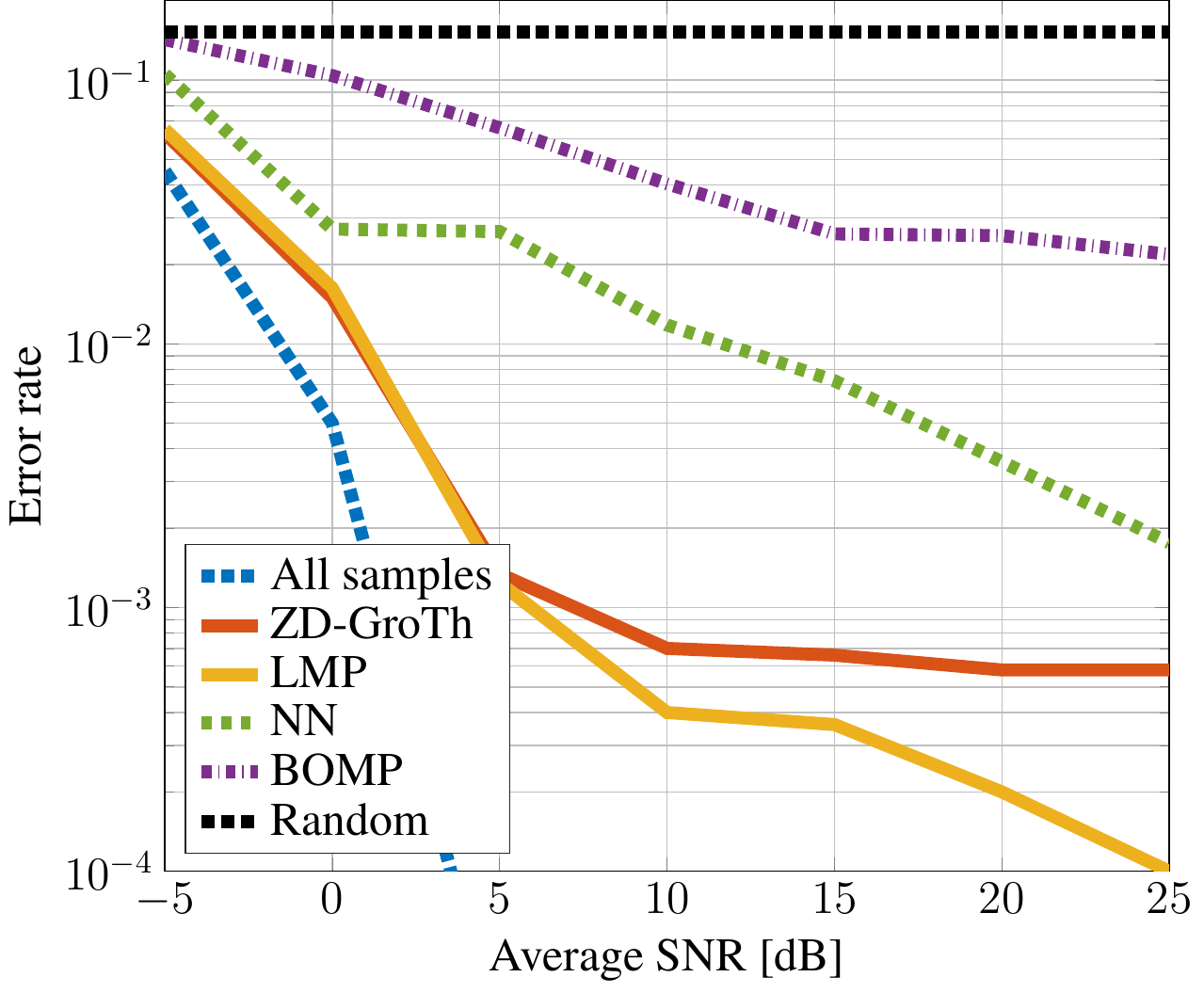}
		\label{fig:100}
	}
	\subfloat[150 NUWS Samples]
	{
		\includegraphics[width=\figsize\textwidth]{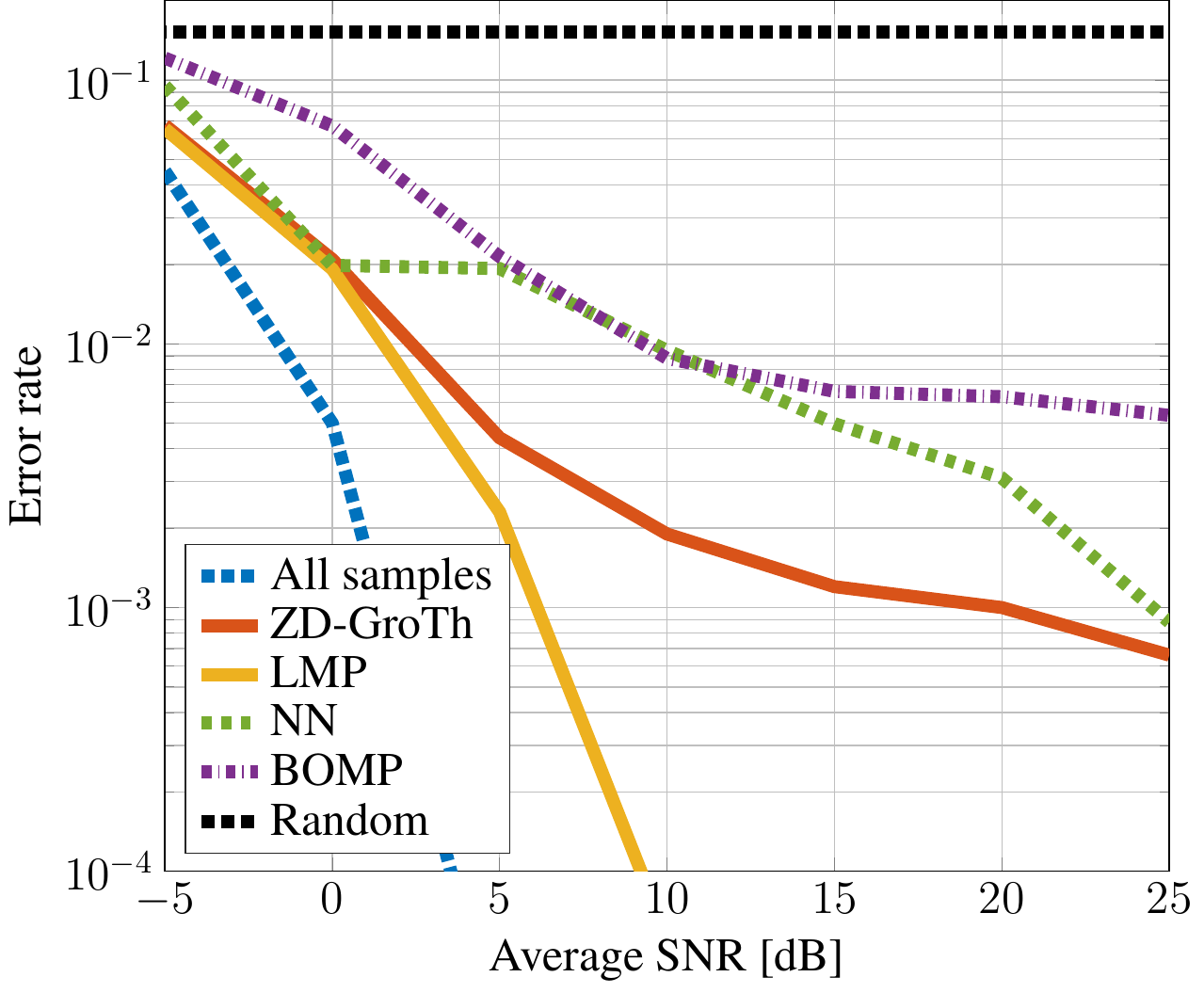}
		\label{fig:150}
	}
	\vspace{0.0cm}
	\caption{Comparison of algorithms that identify unused RF channels: BOMP, ZD-GroTh, LMP, and neural network (NN)-based detector; (a) $ 50 $ NUWS samples; (b) $ 100 $ NUWS samples, and (c) $ 150 $ NUWS samples; as baseline methods, we include the performance of identifying an unused channel using all the Nyquist samples and the performance of random guessing. Clearly, LMP and ZD-GroTh outperform both conventional BOMP and the NN-based approach, whereas LMP has a clear advantage over ZD-GroTh at high SNR by reducing the dynamic range and using an improved correlation criterion.}
	\vspace{-0.345cm}
\end{figure*}

\subsection{Simulation Results}
We show simulation results for $M=50$, $M=100$, and $M=150$ NUWS measurements (corresponding to compression ratios of $1/4$, $1/2$, and $3/4$) that are obtained by performing $50,000$ Monte-Carlo trials, which randomize transmitter location, spectrum occupancy, transmit signals, and noise.
Figures \ref{fig:50}, \ref{fig:100}, and \ref{fig:150} show the resulting error rates vs. average signal-to-noise ratio (SNR) over all active transmitters. 
Errors are declared whenever we decide a channel was unused but the channel was occupied by a transmitter. 
Black dashed lines show the baseline error rate for randomly selecting an unused channel; blue dashed lines show the error rate of using all the $N=200$ Nyquist samples and analyzing the signal power in the discrete Fourier domain. 
Green dashed lines show the error rate of a feedforward neural network (NN) with four hidden layers, each having $ 1024$, $512$, $256$, and $128$ neurons with rectified linear unit (ReLU) activation functions. 
We train the NN from 200,000 examples.
Purple dashed lines show the error-rate of BOMP with $K=19$ iterations which leaves one channel left that we declare as unused and red solid lines show the error rate of ZD-GroTh as in \eqref{eq:antiomp}. Orange solid lines show the error rate  of the proposed LMP algorithm with  $P=4$. 

We observe that both ZD-GroTh and LMP consistently achieve lower error rate than BOMP and the NN-based whitespace detector, even though the neural network has been retrained for each SNR. 
Furthermore, we see that at moderate SNR values ($\textit{SNR}\geq5$\,dB),  LMP outperforms ZD-GroTh. At low SNR   ($\textit{SNR}<5$\,dB),  performing NUWS on 200 time samples is insufficient to achieve low error rates. If lower error rates are desirable, more samples have to be acquired---the same observation applies to the Nyquist-based approach. 

\section{Conclusions}
We have proposed a novel algorithm, called \emph{least matching pursuit} (LMP), for detection of unused blocks from CS measurements. 
The design of LMP is inspired by a novel recovery guarantee designed for the zero-detection group thresholding (ZD-GroTH) method put forward in~\cite{yoo2013finding}. 
In contrast to ZD-GroTH, LMP first eliminates the strongest signal components using block orthogonal matching pursuit (BOMP) to reduce the dynamic range in the residual signal. LMP then evaluates a minimum correlation criterion that uses intermediate results acquired during BOMP iterations.
Simulation results with realistic RF components and a nonuniform wavelet sampling (NUWS) stage have shown that LMP is able to outperform BOMP, ZD-GroTH, and a neural-nework-based baseline detector for compressive whitespace detection.  

While LMP achieves lower error rates that existing methods, its performance can further be improved by using a generalized algorithm for an ``anti'' multiple measurement vectors (MMV) problem~\cite{cotter2005sparse}, which performs averaging over more time slots. A corresponding study is part of ongoing work. 
A theoretical analysis of the proposed average minimum correlation criterion is an open research problem and evaluating the performance of LMP for multipath channels is part of ongoing work.
Finally, we are developing a hardware prototype that performs NUWS and LMP in an energy-efficient manner, which will demonstrate the real-world performance of our approach. 

\appendices

\section{Proof of Proposition~\ref{prop:bomp}}
\label{app:bomp}
A sufficient condition for ZD-GroTh to succeed is 
\begin{align}
& \min_{i\in \mathcal{N}} \| (\bA_i^H\bA_i)^{-0.5}\bA_i^H \bmy\|_2  \notag \\
& \quad \qquad < \min_{j\in \mathcal{U}} \| (\bA_j^H\bA_j)^{-0.5}\bA_j^H \bmy\|_2, \label{eq:cond1}
\end{align}
which holds true if the smallest correlation with an unused block in the set $\setN$ is strictly smaller than the smallest correlation with a used block in $
\setU$. 
The proof follows by upper-bounding and lower-bounding the left-hand side (LHS) and right-hand side (RHS) of~\fref{eq:cond1}, respectively.
An upper bound to the LHS of~\fref{eq:cond1} is as follows:
\begin{align}
& \min_{i\in \mathcal{N}} \| (\bA_i^H\bA_i)^{-0.5}\bA_i^H \bmy\|_2 \\
& \quad = \min_{i\in \mathcal{N}} \left\|(\bA_i^H\bA_i)^{-0.5}\bA_i^H\left(  \sum_{j\in \mathcal{U}} \bA_j \bmx_j+\bmn\right)\right\|_2 \\
 &\quad \leq \mu_B   \sum_{j\in \mathcal{U}} \|\bmx_j\|_2+\|\bmn\|_2. \label{eq:LHSbound}
\end{align}
Here, the inequality follows from the triangle inequality, the definition of the block mutual coherence $\mu_B$, and the fact that our selection criterion is normalized. 
A lower bound on the RHS of~\fref{eq:cond1} is as follows:
\begin{align}
& \min_{j\in \mathcal{U}} \| (\bA_j^H\bA_j)^{-0.5}\bA_j^H \bmy\|_2 \\
&  = \min_{j\in \mathcal{U}}  \left\| (\bA_j^H\bA_j)^{-0.5}\bA_j^H \!\left(\bA_j\bmx_{j} +\!\!\!  \sum_{l\in \mathcal{U}\backslash j	} \bA_l \bmx_l +\bmn\right)\right\|_2 \\
&  \geq \sigma_{\min}\|\bmx_{\min}\|_2- \mu_B  \sum_{j\in \mathcal{U}} \|\bmx_j\|_2 + \mu_B \|\bmx_{\min}\|_2 -\|\bmn\|_2. \label{eq:RHSbound}
\end{align}
Here, the inequality follows from the definitions of $\sigma_{\min}$ and~$\mu_B$, the fact that for one of the used blocks $\|\bmx_j\|=\|\bmx_{\min}\|$, and the normalized selection criterion.
By substituting the LHS and RHS in the sufficient condition \eqref{eq:cond1} by \fref{eq:LHSbound} and~\fref{eq:RHSbound}, respectively, we arrive at the final result in \fref{eq:prop1}. 

\balance

\bibliographystyle{IEEEtran} 
\bibliography{IEEEabrv,spawcbib}

\balance

\end{document}